\documentclass[11pt,oneside]{amsart} 
\usepackage{amscd,amssymb,amsxtra}
\usepackage{mathabx}
\usepackage{comment}
\usepackage{enumerate}
\usepackage{color}
\makeatletter
\let\@secnumfont\bfseries
\def\section{\@startsection{section}{1}%
  \z@{4\linespacing\@plus\linespacing}{\linespacing}%
  {\bfseries\centering}}
\def\introsection{\@startsection{section}{1}%
  \z@{3\linespacing\@plus\linespacing}{\linespacing}%
  {\bfseries\centering}}
\def\subsection{\@startsection{subsection}{2}%
   \z@{1.25\linespacing\@plus.7\linespacing}{.5\linespacing}%
   {\normalfont\bfseries}}
\def\subsectionsinline{\def\subsection{\@startsection{subsection}{2}%
  \z@{1\linespacing\@plus.7\linespacing}{-.5em}%
  {\normalfont\bfseries}}}
\makeatother

\theoremstyle{definition}

\newtheorem*{example*}{Example}
\newtheorem*{problem*}{Problem}
\newtheorem*{exercise*}{Exercise}
\newtheorem*{question*}{Question}

\theoremstyle{remark}

\newtheorem*{note*}{Note}
\newtheorem*{notation*}{Notation}
\newtheorem*{remark*}{Remark}

\theoremstyle{plain}
\newtheorem{definition}[equation]{Definition}

\newtheorem*{definition*}{Definition}
\newtheorem*{theorem*}{Theorem}
\newtheorem*{corollary*}{Corollary}
\newtheorem*{lemma*}{Lemma}
\newtheorem*{proposition*}{Proposition}
\newtheorem*{conjecture*}{Conjecture}
\newtheorem*{claim*}{Claim}
\newtheorem*{proposal*}{Proposal}
\newtheorem*{conclusion*}{Conclusion}

\numberwithin{equation}{section}

\definecolor{refkey}{rgb}{0,.6,.4}

\renewcommand{\:}{\colon}
\newcommand{\Ahat}{{\hat A}}

\newcommand{\CC}{{\mathbb C}}

\newcommand{\EE}{\mathbb E}

\DeclareMathOperator{\pt}{pt}

\newcommand{\RR}{{\mathbb R}}
\newcommand{\TT}{\mathbb T}
\DeclareMathOperator{\Spin}{Spin}

\newcommand{\ZZ}{{\mathbb Z}}
\DeclareMathOperator{\ch}{ch}
\newcommand{\chiup}{\raise.5ex\hbox{$\chi$}}

\newcommand{\inv}{^{-1}}
\newcommand{\mstrut}{^{\vphantom{1*\prime y}}}

\newcommand{\temsquare}{\raise3.5pt\hbox{\boxed{ }}}

\newcommand{\zmod}[1]{\ZZ/#1\ZZ}

\newcommand{\zt}{\zmod2}

\usepackage{hyperref}
\usepackage{graphicx}

\DeclareMathOperator{\codim}{codim}
\DeclareMathOperator{\ob}{ob}
\DeclareMathOperator{\pfaff}{pfaff}

\newcommand{\cH}{\widecheck{H}}
\newcommand{\cKR}{\widecheck{KR}}
\newcommand{\cR}{\widecheck{R}}
\newcommand{\cj}{\check{j}}
\newcommand{\ctau}{\check\tau }
\newcommand{\ct}{\check\beta }
\newcommand{\cz}{\check\zeta }
\newcommand{\gpd}{/\!/\,} 
\newcommand{\hS}{\widehat{\Sigma }}
\newcommand{\hp}{\hat{\pi }}
\newcommand{\hw}{\hat{w}}
\newcommand{\jext}{j_{\textnormal{ext}}}

\begin{document}

\abovedisplayskip18pt plus4.5pt minus9pt
\belowdisplayskip \abovedisplayskip
\abovedisplayshortskip0pt plus4.5pt
\belowdisplayshortskip10.5pt plus4.5pt minus6pt
\baselineskip=15 truept
\marginparwidth=55pt

\renewcommand{\theequation}{\arabic{equation}}
\renewcommand{\labelenumi}{\textnormal{(\roman{enumi})}}




 \title[Orientifold Pr\'ecis]{Orientifold Pr\'ecis} 

 \author[J. Distler]{Jacques Distler} 
 \thanks{The work of J.D. is supported by the National Science Foundation
under grant PHY-0455649}   
 \address{Theory Group, Department of Physics, and Texas Cosmology Center \\
University of Texas \\ 1 University Station C1600\\ Austin, TX 78712-0264}
 \email{distler@golem.ph.utexas.edu}

 \author[D. S. Freed]{Daniel S.~Freed}
 \thanks{The work of D.S.F. is supported by the National Science Foundation
under grant DMS-0603964}
 \address{Department of Mathematics \\ University of Texas \\ 1 University
Station C1200\\ Austin, TX 78712-0257}
 \email{dafr@math.utexas.edu}

 \author[G. W. Moore]{Gregory W.~Moore}
 \thanks{The work of G.W.M. is supported by the DOE under grant
DE-FG02-96ER40949.}
 \address{NHETC and Department of Physics and Astronomy \\
Rutgers University \\ Piscataway, NJ 08855--0849}
 \email{gmoore@physics.rutgers.edu}
 \thanks{We also thank the Aspen Center for Physics for providing a
stimulating environment for many discussions related to this paper.}
 \thanks{Report numbers: UTTG-04-09, TCC-019-09.} 
 \date{\today}
 \begin{abstract} 
 We give a precise and concise formulation of the orientifold construction in
Type II superstring theory.  Our results include anomaly cancellation on the
worldsheet and a $K$-theoretic computation of the background Ramond-Ramond
charge.
 \end{abstract}
\maketitle



Since its inception~\cite{Sa,PS,Ho,DLP,BiS} the orientifold construction of
Type II superstring theory has proven useful both for the formal development
of string theory and for its potential applications to phenomenology.  For
reviews see \cite{P,D,AnS,BH1,J,BH2,BHHW}.  Our work began many years ago
when the first author showed the second the formula~\cite{DJM,CR,MSS,S,SS}
for the Ramond-Ramond charge induced by an orientifold and inquired about its
$K$-theoretic significance, especially in view of the unusual cousin of the
Hirzebruch $L$-genus contained therein.  In the process of our investigations
we were led to more foundational questions about orientifolds and the Type~II
superstring.  In this letter we describe some mathematical foundations we
have developed to resolve these questions.  In particular, we give careful
definitions of both the worldsheet and spacetime fields of an orientifold,
including precise Dirac quantization conditions for the $B$-field and
Ramond-Ramond (RR) field.  The $(\textnormal{Neveu-Schwarz})^2=$NSNS fields
which appear in Definition~\ref{thm:1} involve subtle topological structures
which, for very different reasons, exactly fit what is needed in both the
(short distance) worldsheet and (long distance) spacetime theories.  The
worldsheet fields are enumerated in Definition~\ref{thm:2}.  A surprising
challenge here is to define the integral of the $B$-field over the
worldsheet.  The lack of a proper orientation leads to a novel prescription,
one feature of which is that the $B$-field amplitude has a (classical)
anomaly in the sense that it takes values in a complex line not canonically
isomorphic to the complex numbers.  It then cancels against the more standard
(quantum) anomaly from the spinor field on the worldsheet, and the
cancellation uses the ``twisted'' spin structure in spacetime.\footnote{The
twisted spin structure appears in Definition~\ref{thm:1}.}  The spacetime RR
field is self-dual, so part of its formulation (Definition~\ref{thm:3})
involves a certain quadratic function.  The general theory of self-dual fields
defines an RR charge due to the orientifold background, which we compute with
the prime $2$ inverted, in which case it localizes to the orientifold fixed
point set.  Tensoring with the reals we recover the formula~\eqref{eq:6}
which began this project.

Our definitions are new, even in the case of the Type~I superstring, and
offer some refinements of the standard (non-orientifold) Type~II superstring.
Taken together the orientifold data is an impressively tight structure, and
leads to the most intricate matching we know between topological features in
a short distance theory and its long distance approximation.  We illustrate
the tight flow of ideas in Figure~1: the basic hypotheses are in the three
turquoise ovals on top, the consequences for the worldsheet theory in the
green diamonds, and the consequences for the spacetime theory in the yellow
rectangles.

\begin{figure}
\begin{center}
 \centering\includegraphics[scale=.45]{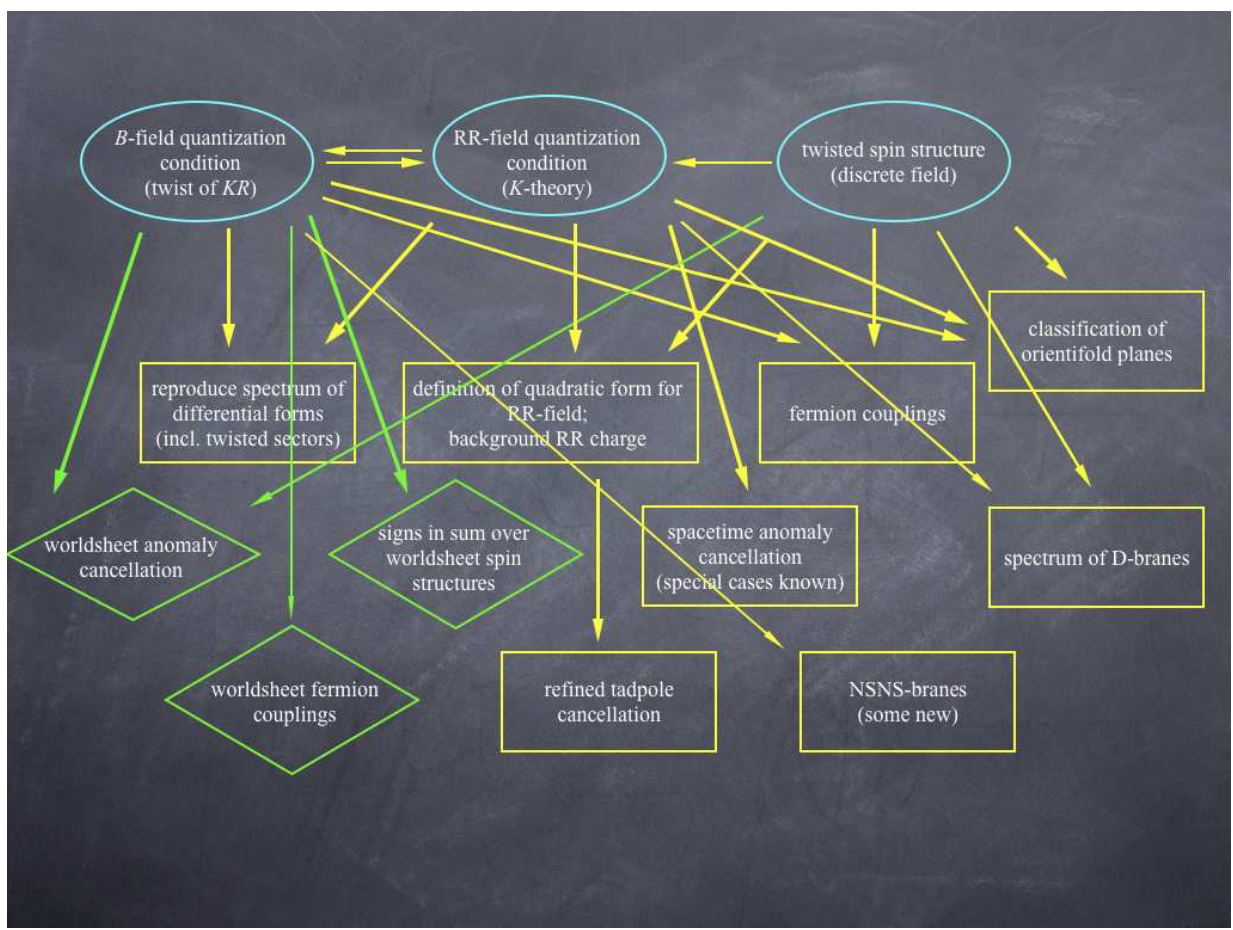}
\end{center} 
\caption{Structure of fields impacts worldsheet and spacetime theories}
 \end{figure}

In recent years new abelian objects have entered differential geometry.  Most
familiar are ``gerbes with connection'' in various incarnations.  Often these
objects have cohomological significance and can be studied as part of
generalized differential cohomology and twisted versions thereof.  The
abelian gauge fields in this paper are examples of such objects.  Conversely,
these geometrical constructs are precisely what we need to formulate
orientifolds.  The foundations of \emph{equivariant} generalized differential
cohomology have yet to be fully developed; see~\cite{SV,O,BSh} for recent
accounts of equivariant differential complex $K$-theory.  On the other hand,
our results about worldsheet anomalies and the RR~background charge are
purely topological and do not require these missing foundations.

We are writing longer accounts (e.g.~\cite{DFM}) which will on the one hand
include detailed definitions, statements, and proofs; and on the other
explain the results in a manner more accessible to physicists, including the
relationship to previous approaches to orientifolds and several physical
consequences.  One of the virtues of Definition~\ref{thm:1} is that it tames
the zoo of orientifolds into a small set of data; in subsequent papers we
will unpack this definition to recreate the zoo.  Also, our formulation of
the $B$-field allows us to state new consistency conditions for D-branes in
orientifolds, and suggests the existence of new NS-branes with torsion
charges.  We have not investigated anomalies in the long distance spacetime
theory, but have provided a framework in which to investigate them.
 
Previous work on $K$-theoretic interpretation of RR~charge in orientifolds
includes~\cite{W2,G,Hor,BGH,OS,dBHKMMS,AH,GL,BGS,BS}.  The twisted
$KR$-theory used here unifies all the various forms of $K$-theory in these
papers.

We thank Michael Hopkins for topological consultations.  We also thank Ilka
Brunner, Frederik Denef, Emanuel Diaconescu, David Morrison, Graeme Segal,
Constantin Teleman, and Edward Witten for useful discussions.

 \subsection*{NSNS superstring backgrounds}\label{subsec:1.1}

We begin with some general concepts.

A \emph{smooth orbifold} of dimension~$n$ is a space locally modeled on the
quotient of~$\RR^n$ by the linear action of a finite group, and it has a
smooth structure defined similarly to that of a smooth manifold;
see~\cite{ALR} for a recent exposition.  (Smooth manifolds are smooth
orbifolds.)  An important example to keep in mind is the case of a global
quotient, where a discrete group $\Gamma$ acts on a smooth manifold~ $Y$ with
finite stabilizers; the quotient smooth orbifold is denoted~$Y\gpd\Gamma $.
Orbifolds are presented as groupoids, which may or may not be global
quotients.  A smooth orbifold can also be termed a `smooth real
Deligne-Mumford stack', and this leads to a more invariant description, but
as that s-word conjures up demons, we avoid it.  A smooth orbifold has a
geometric realization, which in the case of a global quotient~$Y\gpd\Gamma $
is known as the Borel construction.  We define the ordinary cohomology of an
orbifold to be that of the geometric realization.
 
We make use of generalized cohomology theories~\cite{A}, such as periodic
$KO$-theory and its connective cover~$ko$, and also employ twisted versions.
Twistings are geometric objects whose equivalence class can be located in a
cohomology theory, at least if the cohomology theory being twisted is
suitably multiplicative.  The cohomological degree is a particular type of
twisting.  For example, on an orbifold~$X$ the twistings of $K$-theory of
interest here~\cite{FHT} are classified by the \emph{set}
  \begin{equation}\label{eq:7}
     H^0(X;\zt)\times H^1(X;\zt)\times H^3(X;\ZZ). 
  \end{equation}
Let~$R$ be the Postnikov section ~$ko\langle 0\cdots 4 \rangle$ of~$ko$; it
has homotopy groups concentrated in the indicated degrees.  The twistings of
$K$-theory we use are classified by the \emph{abelian group}~$R\inv (X)$,
which as a set is isomorphic to~\eqref{eq:7}.  It is crucial for us that
$R$~is a \emph{multiplicative} cohomology theory, more precisely an
$E^{\infty}$~ring: we can multiply and integrate.  Twistings and orientations
of~$ko$ (or of the more familiar periodic version~$KO$) induce twistings and
orientations of~$R$.
 
Abelian gauge fields and their associated currents are geometric objects
which live in \emph{differential} cohomology theories~\cite{F1,F2,FMS2}.
The differential theory associated to a cohomology theory~$h$ is
denoted~$\check{h}$.  A systematic development of some foundations is given
in~\cite{HS}.  A differential cohomology group is the set of equivalence
classes in a groupoid.\footnote{To obtain \emph{local} objects, we need to
consider higher groupoids, or spaces.}  For example, objects in degree two
differential cohomology may be taken to be circle bundles with connection;
the differential cohomology \emph{group}~$\cH^2(X)$ is the set of equivalence
classes.  We use the notation~`$\ob$' below to emphasize that the fields are
geometric objects in a groupoid, not equivalence classes in a group.  There
are often different models for a given differential cohomology theory, and we
needn't commit ourselves to a particular model.  Twistings of differential
theories are differential twists of the underlying topological theory.
 
With these preliminaries we can define the $(\textnormal{Neveu-Schwarz})^2$
sector of an orientifold background.

        \begin{definition}[]\label{thm:1}
 An \emph{NSNS superstring background} consists of:
 \begin{enumerate}
 \item a 10-dimensional smooth orbifold~$X$ together with Riemannian metric
and real-valued scalar (dilaton) field; 

 \item a double cover $\pi \:X_w\to X$;

 \item a differential twisting~$\ct$, the \emph{B-field};

 \item and a \emph{twisted spin structure} $\kappa\:\Re(\beta )\to\tau
^{KO}(TX-10)$.

 \end{enumerate}
        \end{definition}

\noindent
 We call $\pi \:X_w\to X$ the \emph{orientifold double cover}.  The
symbol~`$w$' is used to denote the orientifold double cover as well as its
equivalence class in~$H^1(X;\zt)$.  In the case of a global
quotient~$X=Y\gpd\Gamma $, the double cover may be described by an index two
subgroup~$\Gamma _0\subset \Gamma $: then $X_w = Y\gpd\Gamma_0$.  In physics,
one speaks of ``the orientifold of the spacetime $Y$ by the action of
$\Gamma$.''  Our definition includes the usual orbifold construction in
superstring theory.  It also includes non-orientifold theories by positing a
trivialization of $X_w\to X$.  Underlying the $B$-field~$\ct$ is a
twisting~$\beta$ of~$KR(X_w)$ whose equivalence class lies in~$R^{w-1}(X)$,
which is the twisting of~$R\inv (X)$ induced by the orientifold double cover.
As a set $R^{w-1}(X)$~is isomorphic to the product of cohomology
groups~\eqref{eq:7}, but with the last factor replaced by ~$H^3(X;\ZZ_w)$ for
$\ZZ_w$~the local coefficient system associated to~$X_w\to X$.  The
``curvature'' of~$\ct$ is a closed twisted 3-form with integral periods, the
3-form field strength of the $B$-field.  It lifts to an ordinary 3-form
on~$X_w$ which is odd under the orientifold involution on~$X_w$.  What is new
in our definition of the $B$-field, even in non-orientifold theories, is the
inclusion of the $H^0$~and $H^1$~components in~\eqref{eq:7}.  The twisted
spin structure in~(iv) is an isomorphism of twistings of $KO(X)$: $\Re(\beta
)$~is a lift of $\bar\beta +\beta $ to a $KO$-twisting and $\tau
^{KO}(V)$~denotes the $KO$-twisting determined by the real vector bundle~$V$.
In more down-to-earth terms, an ordinary spin structure on an $n$-dimensional
Riemannian manifold (or orbifold) is a reduction of the principal
$O_n$-bundle of orthonormal frames to a $\Spin_n$-bundle.  Note that
$\Spin_n$ is a double cover of an index two subgroup of~$O_n$.  The twisted
spin structure in~(iv) is a similar reduction, but the double cover of the
index two subgroup depends on the topological object~$\beta $ underlying the
$B$-field.  A twisted spin structure is a \emph{discrete field} in the long
distance supergravity theory; it has no differential form field strength.

The $B$-field is classified topologically by~$R\inv (X)$ as in~\eqref{eq:7}.
The $H^0$~component~$t$ distinguishes between the usual Type~B ($t=0$) and
Type~A ~($t=1$).  Its inclusion in the $B$-field incorporates the signs in
the sum over spin structures of the worldsheet Type~II theory~\cite{SW,AgMV}
as part of the $B$-field amplitude~\eqref{eq:3}, as we explain in~\cite{DFM}.
Note that since $R^{-1}(X)$ is a group there is a distinguished zero, thus
singling out the IIB superstring as ``more fundamental.''  The
$H^1$~component~$a$ is a further twisting of a Type~II theory.  We remark
that the existence of a twisted spin structure implies the following
relations between the Stiefel-Whitney classes of spacetime~$X$ and the
$B$-field:
  \begin{equation}\label{eq:2}
     \begin{split} w_1(X) & = t w \\ w_2(X) & = t w^2 + a w .\\ \end{split} 
  \end{equation}

The Type~I superstring on a spin 10-manifold~$Y$ is a special case of
Definition~\ref{thm:1}.  Then $X_w=Y$ with trivial involution and
quotient~$X$ is a ``non-effective'' orbifold, presented as the groupoid
$Y\times \pt\gpd(\zt)$.  The $B$-field reduces to a discrete field in~$\ob
H^2(Y;\zt)$---in particular, both $t$~and $a$~vanish ---and the twisted spin
structure is an ordinary spin structure on~$Y$.  The RR~fields may be modeled
as principal $\Spin_{32}$-bundles ``without vector structure''~\cite{LMST,
W3}.  As another special case, when $X_w=\EE^{10}$ is flat Euclidean space
and $X$~the quotient by a reflection in $9-p$~directions, then the resulting
constraint among~$t$, $a$, and~$p$ reproduces the standard list of consistent
orientifold projections~\cite{P}.

 \subsection*{Worldsheet theory}

We now define a worldsheet in a given NSNS background (in the NSR formalism).
Curiously, while low genus surfaces have been extensively investigated in the
physics literature, a formulation valid for general worldsheets does not seem
to be available, even for the Type~I superstring.

        \begin{definition}[]\label{thm:2}
 Fix an NSNS superstring background as in Definition~\ref{thm:1}.  Then a
\emph{worldsheet} consists of
 \begin{enumerate}
 \item a compact smooth 2-manifold~$\Sigma $ (possibly with boundary) with
Riemannian structure;

 \item a spin structure~$\alpha $ on the orientation double cover
$\hp\:\hS\to\Sigma $ whose underlying orientation is that of~$\hS$; 

 \item a smooth map $\phi \:\Sigma \to X$; 

 \item an isomorphism $\phi ^*w\to\hw$, or equivalently a lift of~$\phi $ to
an equivariant map $\hS\to X_w$;
 
 \item a positive chirality spinor field~$\psi $ on~$\hS$ with
coefficients in ~$\hp^*\phi ^*(TX)$;
 
 \item and a negative chirality spinor field~$\chi $ on~$\hS$ with
coefficients in~$T^*\hS$ (the gravitino).

 \end{enumerate}
        \end{definition}

\noindent
 The orientation double cover~$\hS$ carries a canonical orientation, used
in~(ii), whereas no orientation is assumed on~$\Sigma $, which indeed may be
nonorientable.  We use~`$\hw$' to denote the orientation double cover
of~$\Sigma $.  The spin structure~$\alpha $, which is a discrete field
on~$\Sigma $, is locally on~$\Sigma $ a choice of two spin structures with
opposite orientations.  The isomorphism in~(iv) is also a discrete field
on~$\Sigma $.  In case $X=Y\gpd\Gamma $~is a global quotient with double
cover~$X_w=Y\gpd\Gamma _0$, a map~(iii) is given by a principal $\Gamma
$-bundle $P\to \Sigma $ and a $\Gamma $-equivariant map $P\to Y$.
Furthermore, $P$~is oriented, elements of~$\Gamma _0$~preserve the
orientation, and elements of~$\Gamma \setminus\Gamma _0$ reverse it.  The
spinor fields in~(v) and~(vi) use the spin structure~$\alpha $; the chirality
refers to the canonical orientation of~$\hS$.
 
Assume the boundary of~$\Sigma $ is empty.  The exponentiated Euclidean
action, after integrating out~$\psi $ and~$\chi $, has two factors on which
we focus:
  \begin{equation}\label{eq:3}
     \exp\Bigl(2\pi i\int_{\Sigma }\cz\cdot\phi ^*\ct\Bigr)\cdot \pfaff
     D_{\hS,\alpha  }\bigl(\hp^*\phi ^*(TX)-T\Sigma \bigr),  
  \end{equation}
the $B$-field amplitude and a pfaffian.  In the first factor the pullback
$\phi ^*\ct$ is an object in~$\cR^{\hw-1}(\Sigma )$, because of~(iv) in
Definition~\ref{thm:2}.  Sadly, $\Sigma $~is not endowed with an
$R$-orientation\footnote{An $R$-orientation is the same as an
$\cR$-orientation.  Because of the $\hw$~twisting in the integrand we do not
quite need an $R$-orientation, which is a spin structure on~$\Sigma $, but
rather a pin structure.  Still, we do not have one.} and so we have
introduced a new object~$\cz$ in twisted differential $R$-theory in order to
define the integral.  We do not give here a precise definition, but remark
that one ingredient is a pushforward of~$\alpha $ to~$\Sigma $, which
measures the obstruction to refining~$\alpha $ to a pin structure.  Of utmost
importance is that the first factor in~\eqref{eq:3} is not a number but
rather an element in a certain hermitian ``$B$-line''~$L_B$: the $B$-field
amplitude is anomalous.  Let~$S$ be a parameter space of non-fermionic
worldsheet fields (Definition~\ref{thm:2}(i)--(iv)).  In string theory one
integrates over worldsheets, and therefore the exponentiated effective action
after integrating out the fermions should be a \emph{measure}
on~$S$.\footnote{In this letter we ignore all subtleties associated with the
measure for supermoduli. We believe these are unrelated to the questions we
address here.}  The measure turns out to be a product of a measure which is
manifestly well-defined on~$S$ and ~\eqref{eq:3}, whence the latter should be
a function on~$S$.  The pfaffian of the Dirac operator on~$\hS$ has its usual
anomaly: it takes values in a hermitian ``pfaffian line''~$L_{\psi }$.  In a
parametrized family both~$L_B$ and~$L_\psi $ are \emph{flat} line bundles
over~$S$: these are global anomalies.  The crucial result is that the tensor
product~$L_B\otimes L_\psi \to S$ has a \emph{natural} trivialization, and so
defines~\eqref{eq:3} as a function on~$S$.  This trivialization uses the
twisted spin structure on spacetime~$X$ (see Definition~\ref{thm:1}(iv)); it
is the manner in which that twisted spin structure enters the lagrangian
worldsheet theory.  This anomaly cancellation---more precisely, this
``setting of the quantum integrand''~\cite{FM}---is the most subtle example
of its kind that we've seen.  It applies as well in the non-orientifold case:
then the $B$-field amplitude is a function, as the worldsheet carries a spin
structure, and the trivialization of the pfaffian line bundle $L_\psi \to S$
depends on the spin structure on spacetime.
 
We do not attempt to couple the worldsheet to the RR field or fermions on
spacetime.

 \subsection*{The RR field on spacetime}

The RR field is self-dual, so we begin with some general remarks about
self-dual fields.  First, the cohomology theory used to quantize the
self-dual charges and fluxes must itself be Pontrjagin
self-dual~\cite[Appendix~B]{FMS1}.  Furthermore, part of the definition of a
self-dual field is a quadratic refinement of the pairing between electric and
magnetic currents: this refinement is a topological datum.  This quadratic
function has a well-defined center of symmetry~$\mu $.  We interpret~$-\mu $ as
the self-dual charge induced by the background.  If $\mu $~ is nonzero and
$X$~is compact, then there must be additional charged objects---D-branes---in
the theory whose total charge~$\jext$ equals~$\mu $.  These charges have
differential refinements\footnote{The lift of~$\mu $ to a current may involve
an additional topological choice.}---the currents---and the self-dual gauge
field is a (nonflat) isomorphism $\check\mu \to\cj_{\textnormal{ext}}$.  See
\cite{W1,F1,F2,HS,FMS1,BM} for background on self-dual fields.
 
Fix an NSNS superstring background.  The RR charges and fluxes are quantized
by a twisted form of periodic $K$-theory on~$X$: the $KR$~theory on~$X_w$
with its involution.\footnote{Objects in~$KR(X_w)$ carry a lift of the
involution on~$X_w$, so descend to~$X$ and should be regarded as living
on~$X$.  But there is no standard notation for $KR(X_w)$ as a twisted
$K$-group on~$X$.}  As $X_w$~is an orbifold, not a manifold, we must specify
what we mean by its $KR$-theory.  Here we do not use the geometric
realization or Borel construction, but rather use a geometric model which
generalizes the equivariant vector bundles of Atiyah-Segal equivariant
$K$-theory.  (The analogous model for complex $K$-theory appears in
~\cite[\S3]{FHT}.)  The quadratic function is a single topological choice
which induces quadratic functions on families of manifolds of dimension
~$\le12$. Its simplest manifestation is integer-valued and occurs on a
12-dimensional orbifold~$M$ which has a double cover~$M_w$, a
$B$-field~$\ct$, and a twisted spin structure~$\kappa $.\footnote{The twisted
spin structure on~$M$ is an isomorphism $\kappa \:\Re(\beta )\to\tau
^{KO}(TM-12)$.}  

        \begin{definition}[]\label{thm:3}
 Fix an NSNS superstring background as in Definition~\ref{thm:1}.  Then
 \begin{enumerate}
 \item an \emph{RR current} is an object in $\cKR^{\,\ct}\!(X_w)$; 

 \item the required quadratic function on a 12-manifold~$M$ is the composition 
  \begin{equation}\label{eq:4}
     \begin{aligned} KR^{\beta}(M_w)&\longrightarrow KO^{\Re(\beta
      )}_{\zt}(M_w)&\cong \quad  KO^{\tau
      ^{KO}(TM-4)}_{\zt}(M_w)&\longrightarrow KO_{\zt}^{-4}(\pt)\longrightarrow
      \ZZ \\  
       j \qquad &\longmapsto  & \kappa \overline{j}j\qquad\qquad \qquad \qquad
     \quad\;  &\longmapsto \quad 
      \int_{M_w}\kappa \overline{j}j \quad\, \longmapsto \textnormal{$\epsilon
     $-component} 
  \end{aligned} 
  \end{equation}
 \end{enumerate}
        \end{definition}

\noindent
 Notice that the $B$-field~$\ct$ is used in the definition of the RR current
to twist differential $KR$-theory.  The twisted spin structure enters into
the definition of the quadratic function at the second stage, and we have
used Bott periodicity to adjust the degree.  At the last stage of the
quadratic function we identify the quaternionic representation
group~$KO_{\zt}^{-4}(\pt)$ of~$\zt$ with~$\ZZ\oplus \ZZ\epsilon$, where
$\epsilon $~is the sign representation.  Our notation is schematic; details
will appear in a subsequent paper.  The quadratic function~\eqref{eq:4}
generalizes that for non-orientifold Type~II~\cite{W4,FH1,DMW} and for
Type~I~\cite{MW,F1}.
 
One of our main results is the computation of the center~$\mu $ of the
quadratic function~\eqref{eq:4}---which equals \emph{minus} the RR charge of the
orientifold background---but only after inverting~$2$.  Assume $X_w$~is a
10-\emph{manifold} with involution~$\sigma $.  Let~$i\:F\hookrightarrow X_w$
be the fixed point set of the involution and $\nu \to F$~the normal bundle.
We apply the localization theorem in $\zt$-equivariant $KO$-theory~\cite{AS}
to compute the image of~$\mu $ in a certain localization of~$KR^\beta (X)$.
The formula generalizes that in~\cite{FH2} for the special case of the Type~I
superstring.  There a $KO$-theory analog of the Wu class appears, and in the
general story it appears in a twisted form.  Let $r=\codim\mstrut _{X_w}(F)$.
The RR~charge in (possibly twisted) rational cohomology, obtained as a
normalized\footnote{The normalization by~$\sqrt{\Ahat(X)}$ is discussed
in~\cite{GHM,CY,MM,F1}.}  Chern character, is
  \begin{equation}\label{eq:6}
     -\,\sqrt{\Ahat(X)}\ch(\mu )=\pm \,2^{5-r} i_*\left(
     \sqrt{\frac{L'(F)}{L'(\nu )}}\right),\qquad L'(V)=
     \prod\limits_{}\frac{x/4u}{\tanh x/4u},
  \end{equation}
where the $L'$-genus of a real vector bundle~$V$ is expressed as usual in
terms of formal degree two classes, $\Ahat$~is the A-hat genus, and $u$~is
the Bott generator of~$K^2(\pt)$.  The $L'$-genus is reminiscent of
Hirzebruch's $L$-genus, but the factors of~$4$ in the $L'$-genus are not seen
in ordinary index theory.  The sign in~\eqref{eq:6} (as well as omitted
powers of the Bott elements in $K$ and~$KR$) depends on the twisting~$\beta
$.  We remark that under the usual definition of $O^{\pm}$-planes the sign
in~\eqref{eq:6} is~$\pm$.  We will give a precise formula for the sign, as
well as a proof of~\eqref{eq:6} and its $K$-theory progenitor, in a
subsequent paper.

 \subsection*{Physical remarks}

We conclude with some points where our work illuminates the physics.

\begin{enumerate}[(a)]

\item In the physics literature on global quotients one finds different
species of ``orientifold planes''.  Our work gives an intrinsic definition as
well as global constraints on the distribution of orientifold planes.  (Some
examples of these constraints for toroidal orientifolds were investigated in
\cite{dBHKMMS,BGS,GH}.)

\item Our formula~\eqref{eq:6} for the RR charge induced by the orientifold
is compatible with \cite{MSS,SS} but not with
formul\ae\ in \cite{MS,HJ}.

\item In the spacetime supergravity action, there are couplings between the
spacetime fermions and the RR fields. Our definition of the twisted spin
structure (for the fermions) and the twisting of differential $KR$-theory
(for the RR fields) are such as to allow those couplings to be globally
consistently defined in an orientifold background.

\item The case $w=0$ corresponds to Type II string theory.  In this case a
twisted spin structure determines a spin structure on $X$. Moreover, the data
$(t,a)$, which correspond to a graded real line bundle on $X$, can be used to
produce a modified spin structure. One physical interpretation of this pair
of spin structures is that they are the spin structures ``seen'' by the two
gravitinos. (When $t=0$ the two spin structures have the same underlying
orientation, and when $t=1$ they have opposite orientation.) Thus,
introducing twistings with nonzero $a$ incorporates and generalizes
Scherk-Schwarz compactifications~\cite{SSc1,SSc2,KM,H}.

\item For orbifolds which are global quotients~$X=Y\gpd\Gamma $ there is a
subgroup of discrete $B$-fields classified by $H^3_{\Gamma}(\pt;\ZZ)$; these
are known as ``discrete torsion''.  There has been some controversy and
confusion in the literature concerning the generalization to orientifolds.
In our formulation the answer is clear: discrete torsion is classified by
$R^{w-1}_{\Gamma}(\pt)$.  Whereas $H^3_{\Gamma }(\pt;\ZZ)$ classifies central
extensions of~$\Gamma $ by the circle group~$\TT$, the group $R^{w-1}_{\Gamma
}(\pt)$ classifies non-central \emph{$\zt$-graded} extensions of~$\Gamma $
by~$\TT$, the action of $\Gamma $ on~$\TT$ being determined by the
orientifold.  (See also \cite{BS}.)

\item In the case that spacetime is a global quotient $Y\gpd\Gamma$ there is
a model for $KR^0(X_w)$ which makes contact with the tachyon field picture of
$K$-theoretic charges \cite{W2} and is a slight generalization of both standard
equivariant $K$-theory and $KR$-theory.  The Chan-Paton bundle of the
unstable brane filling spacetime---an object in~$KR^0(X_w)$---is a
$\zt$-graded complex vector bundle with the $\Gamma$~action on~$Y$ lifted:
elements of~ $\Gamma_0$ act $\CC$-linearly and elements of $\Gamma-\Gamma_0$
act $\CC$-\emph{antilinearly}.  The tachyon field is an odd endomorphism
graded commuting with this $\Gamma$ action.

\item Suppose $W\subset X$ is the worldvolume of a D-brane.  The open string
field configurations on the D-brane include an object in differential
$KR$-theory on~$W$ with twisting~$\ctau\mstrut _{W}$.  The induced RR~current
is computed by pushing forward under the inclusion~$i\:W\hookrightarrow X$,
from which we deduce an important constraint relating the twisting class
on~$W$, the topology of $W$, and the $B$-field: namely, there must exist an
isomorphism~$\tau\mstrut _W + \tau^{KR}(\nu) \cong i^*\beta$
in~$R^{i^*w-1}(W)$, where $\nu \to W$ is the normal bundle.  In the
non-orientifold case this is the spacetime derivation of the anomaly derived
in~\cite{FW} from the open string worldsheet.  In case $W$~ is spin and
coincides with an orientifold plane, we easily recover the standard rules for
when a D-brane supports an orthogonal or symplectic gauge theory.  There are
many more possibilities in general.  The need to view D-brane sources in
orientifolds in a $K$-theoretic way in order to avoid paradoxes is central to
the work of \cite{CDE}.

\item Our work raises the possibility that there exist new solitonic objects
in superstring theory. The magnetic NS~current is an object in $\cR^{w+0}(X)$
while the electric NS~current is an object in $\cR^{w+8}(X)$. If we consider
NS-charged branes with constant charge density on their worldvolume, then the
$R$-cohomology classification of NS~currents recovers the fundamental string
and the solitonic 5-brane.  It also predicts the existence of an electrically
charged particle with $\zt$ torsion charge, as well as magnetically charged
$\zt$-torsion 7- and 8-branes. The 8-brane seems to be especially curious,
being a domain wall between the type IIA and IIB theories.

\item It would be interesting to reconcile our Dirac charge quantization
condition for the $B$-field---which is different from that of the bosonic
string---with that of the pure spinor formalism~\cite{B}.

\item We hope that our formulation of orientifold theory can help clarify
some aspects of and prove useful to investigations in  orientifold
compactifications, especially in the applications to model building
and the ``landscape.''  In particular, our work suggests the existence of
topological constraints on orientifold compactifications which have not been
accounted for in the existing literature on the landscape.
\end{enumerate}

\newcommand{\etalchar}[1]{$^{#1}$}
\providecommand{\bysame}{\leavevmode\hbox to3em{\hrulefill}\thinspace}
\providecommand{\MR}{\relax\ifhmode\unskip\space\fi MR }
\providecommand{\MRhref}[2]{%
  \href{http://www.ams.org/mathscinet-getitem?mr=#1}{#2}
}
\providecommand{\href}[2]{#2}

\end{document}